# Observation of flat bands due to band hybridization in $3d$-electron heavy-fermion compound CaCu$_3$Ru$_4$O$_{12}$

Haijiang Liu,[1,2] Yingying Cao,[1,2] Yuanji Xu,[1,2] D. J. Gawryluk,[3,*] E. Pomjakushina,[3] S.-Y. Gao,[1,2] Pavel Dudin,[4] M. Shi,[5] Yi-feng Yang,[1,2,6,†] and H. Ding[1,2,6,†]

*[1]Beijing National Laboratory for Condensed Matter Physics and Institute of Physics, Chinese Academy of Sciences, Beijing 100190, China*
*[2]University of Chinese Academy of Sciences, Beijing 100049, China*
*[3]Laboratory for Multiscale Materials Experiments, Paul Scherrer Institute, Villigen CH-5232, Switzerland*
*[4]Diamond Light Source, Harwell Science and Innovation Campus, Didcot OX11 0DE, United Kingdom*
*[5]Swiss Light Source, Paul Scherrer Institute, CH-5232 Villigen PSI, Switzerland*
*[6]Songshan Lake Materials Laboratory, Dongguan, Guangdong 523808, China*

## ABSTRACT

We report angle-resolved photoemission spectroscopy and first-principles numerical calculations for the band structure evolution of the $3d$ heavy-fermion compound CaCu$_3$Ru$_4$O$_{12}$. Below 200 K, we observed an emergent hybridization gap between the Cu $3d$ electron-like band and the Ru $4d$ hole-like band and the resulting flat band features near the Fermi energy centered around the Brillouin zone corner. Our results confirm the non-Kondo nature of CaCu$_3$Ru$_4$O$_{12}$, in which the Cu $3d_{xy}$ electrons are less correlated and not in the Kondo limit. Comparison between theory and experiment also suggests that other mechanism such as nonlocal interactions or spin fluctuations beyond the local dynamical mean-field theory may be needed in order to give a quantitative explanation of the peculiar properties in this material.

Corresponding author:
† yifeng@iphy.ac.cn , dingh@iphy.ac.cn

* On leave from Institute of Physics, Polish Academy of Sciences, Aleja Lotnikow 32/46, Warsaw PL-02-668, Poland

**INTRODUCTION:**

Heavy-fermion materials are mostly intermetallics with partially filled $4f$ or $5f$ orbitals[1], except for a few examples of $3d$-electron systems that include $LiV_2O_4$, $FeGe_3Te_2$, $CaCu_3Ir_4O_{12}$, etc[2–13]. The mechanism underlying these $3d$ heavy-fermion systems remains unclear. Among them, the $A$-site ordered perovskite $ACu_3B_4O_{12}$ ($A$ = Ca, La, Na; $B$ = Ru, Ir) provides a highly tunable platform for investigating the heavy-fermion properties and mechanism with the perovskite-type structure. $CaCu_3Ru_4O_{12}$ (CCRO) is the earliest member studied in this family. It has an electronic specific heat coefficient of about 84 mJ/(mol $K^2$)[8], which is much larger than that of a normal metal and was initially attributed to the Kondo physics of the Cu localized $3d$ moments. Indeed a broad-maximum feature was observed in the magnetic susceptibility around $T \sim 180 - 200$ K, which was originally identified as the Kondo temperature. Photoemission data[12] also suggested that CCRO is a Kondo system albeit without momentum resolved data. However, a later nuclear magnetic resonance (NMR) experiment up to 700 K observed no signatures of the local moment[11], thus posing a severe challenge to the Kondo scenario. The broad-maximum feature was then explained as a consequence of a peak in the density of states (DOS) just above the Fermi level[13]. So far no decisive conclusions have been reached on the mechanism of these unusual properties, in particular, their correspondence with the electronic band structures on the microscopic level.

In this work, we performed a systematic electronic structure study on CCRO by using angle-resolved photoemission spectroscopy (ARPES) and density functional theory (DFT) calculations in combination with the dynamical mean-field theory (DMFT) for local Coulomb interactions of the Cu 3d orbitals. By tracking carefully the temperature evolution of low-energy electronic structure, we are able to reveal two nearly flat bands near the Fermi energy emerging below 200 K as a result of hybridization between the Cu $3d$ electron-like band and the Ru $4d$ hole-like band. At higher temperatures above 200 K, the hybridization diminished, but the Cu $3d_{xy}$ band still remains near the Fermi energy, which is in strong contrast to the usual Kondo picture and indicates the more itinerant and non-Kondo nature of the Cu 3d electrons in this compound. Our detailed DFT+DMFT calculations confirm the observed band structures but predict a crossover at a much higher temperature, implying the presence of other possible mechanisms such as nonlocal interactions or spin fluctuations beyond local correlation. Comparison between experiment and theory suggests a two-band correlated model, in which the Cu $3d_{xy}$ electrons are not in the Kondo limit. These results provide important information for clarifying the origin of the heavy-fermion properties in CCRO and may also help our understanding of other members of the same family.

**RESULTS AND DISCUSSIONS:**

High-quality single crystals of CCRO were grown by the flux method and cleaved in situ along the (001) plane at $T$ = 15 K with the vacuum below $5 \times 10^{-11}$ mbar. As shown in Fig. 1(a), CCRO adopts the $ABO_3$ perovskite-type crystal structure. The Ca and Cu ions are located at the A site surrounded by 12 nearest oxygen ions. The Ru ions occupy the B site surrounded by 6 nearest oxygen ions forming an octahedron. The first Brillouin zone (BZ) is shown in Fig. 1(b) with the length of Γ-H of 0.85 $Å^{-1}$. The Fermi surface in the (001) plane is same as that of the (100) and (010) planes due to the high symmetry of the lattice structure. ARPES measurements were carried out at the SIS beamline of Swiss Light Source using a SCIENTA R4000 analyzer and at the I05 beamline at the Diamond light source using a SCIENTA R4000 analyzer. Fig. 1(c) displays the wide range valence band spectrum obtained with photon energy $hv$ = 122 eV. Three main features can be observed around -5, -2.6, and -0.1 eV, and are attributed to the O $2p$, Cu $3d$ and Ru $4d$ states, as they are consistent with a previous bulk PES experiment[9] and numerical calculations[10].

To reveal the highly three-dimensional band structure of CCRO, we performed a photon energy dependent measurement with photon energies from 25 eV to 161 eV in steps of 2 eV with the circular polarization along a single cut along the $\Gamma$ - H direction at 15 K, as shown in Fig. 2(a). From the obvious periodicity of the Fermi surface and the cut along the $k_z$ direction, we estimate the inner potential of $V_0 = 13$ eV. From comparison with the calculations, we assign the intense point in Fig. 2(a) to the H point and the faint point to the $\Gamma$ point. Correspondingly, the photon energies crossing the $\Gamma$ or H point are 61, 95, and 122 eV, as shown in Fig. 2(a). The three cuts cross the high symmetry $\Gamma$ or H point along the $\Gamma$ - H path measured at corresponding photon energies, and as shown in Figs. 2(b - d), clearly reveal a flat band feature near the Fermi energy with high intensity centered at the H point, which could be the origin for heavy electrons in this material. In order to show the clear Fermi surface in $\Gamma$ - H - N plane, we display the Fermi surface measured with the photon energy $hv = 122$ eV at T = 6 K with the circular polarization in Fig. 2(e). The Fermi surface map shows obvious strong intensity centered at the H point.

To better understand the cause of this flat band, we carried out temperature dependent measurements along the $\Gamma$ - H direction. We chose the photon energy $hv = 122$ eV which has a high contrast at the H point. As mentioned above, previous magnetic susceptibility measurements show a broad-maximum feature around $T \sim 200$ K, which is the crossover temperature. Figures. 3(a - e) plot the symmetric spectra from 270 to 52 K covering the crossover regime. At 270 K, one can see a hole-like band and an electron-like band centered at the H point. Comparison with first-principles calculations suggests that the hole-like band is composed of the Ru $4d$ orbital, while the electron-like band comes mainly from the Cu $3d$ orbital. With decreasing temperature across 200K, the spectral intensity at the bottom of Cu $3d$ electron band becomes stronger and starts to hybridize with the Ru $4d$ hole band, forming an M-shape hybridized band below the Fermi energy. In Fig. 3(g), the energy distribution curve (EDC) at the H point at $T = 52$ K shows two peaks derived from the upper and lower hybridization bands. From the temperature dependent momentum distribution curves (MDCs) in Fig. 3(h), the peaks reside at around $\pm 0.15$ Å$^{-1}$ at high temperature shrink, indicating the momentum change due to the hybridization between Ru $4d$ and Cu $3d$ electrons.

More information of the band hybridization can be obtained by dividing the measured spectra with the resolution-convoluted Fermi-Dirac function. As shown in Fig. 4, the Ru $4d$ hole pocket and Cu $3d$ electron pocket seem to cross the Fermi level without hybridization at high temperatures above 200 K. When the sample is cooled down, the two bands start to hybridize, forming a M-shape feature around the H point. At 52 K, both of the hybridized bands are intensive with little dispersion around the H point. The EDCs in Fig. 4(g) are chosen at $k = 0$, where we can clearly identify two peaks at 52 K around -68 and 22 meV, showing a hybridization gap of about 90 meV. The EDCs at $k \approx 0.15$ Å$^{-1}$ in Fig. 4(h) shows similar features.

To understand the origin of the band hybridization, we carried out first-principles calculations combining DFT and DMFT[16–19]. The DFT part was computed using the full-potential linearized augmented-plane-wave method implemented in the WIEN2k package[14,15]. The maximum entropy method was used for analytic continuation[21]. We considered non-spin polarized calculations for CCRO with the lattice parameter $a = 7.4082$ Å. The O atomic position is set to (0, 0.1782, 0.3053) with the muffin-tin radii ($R_{MT}$) of 1.70 a.u. for O, 1.99 a.u. for Ru, 1.97 a.u. for Cu and 2.46 a.u. for Ca. We took 2000 **k** points in the Brillouin zone and $R_{MT}K_{max} = 8.0$. The electron correlation of the Cu $3d$ orbitals was treated within the DMFT using the continuous-time quantum Monte Carlo (CTQMC) as the impurity solver[20].

Figure 5(a) plots the DFT+DMFT results at 50 K in comparison with the DFT bands. Near the Fermi energy, we find that only the $d_{xy}$ orbital plays a role among all Cu $3d$ orbitals. A flat (weakly dispersive) band is clearly seen near the H point below the Fermi energy. From the analysis of its orbital character, we conclude that they are the hybridization bands between the more dispersive Ru $4d$ orbitals and the Cu $3d_{xy}$ orbital. These results are consistent with our ARPES data, which reveals the flat band below the Fermi energy at low temperatures. Compared to the DFT results (the solid line), the flat band is more strongly renormalized towards the Fermi energy, suggesting the important role of electronic correlations. It is located at around 300 meV and moves towards the Fermi energy with increasing $U$. However, a very large $U$ beyond 12 eV is needed in order to make comparison with experiment, suggesting the possibility of nonlocal correlations or strong spin fluctuations that may help to enhance the band renormalization in real materials[13].

To see the temperature evolution of the band structures near the Fermi energy more clearly, we compare in Figs. 5(b) and (c) the results at 50 and 1000 K. However, we can only obtain the crossover at a much higher temperature beyond 200 K. This indicates again the possible presence of other interactions beyond DMFT. At high temperatures, the flat band becomes indiscernible. From the spectral function, it appears that the Cu $3d_{xy}$ orbital becomes very broad due to a large imaginary part of its self-energy. However, unlike the usual Kondo lattice systems, the $3d_{xy}$ band stays around the Fermi energy, as observed in ARPES. This indicates that the Cu $3d_{xy}$ electrons are not fully localized in this compound even at high temperatures and the compound is not in the Kondo limit, in contrast to the usual Ce-based heavy fermion compounds. The suppression of the heavy flat band is therefore a result of diminishing hybridization as the $3d_{xy}$ electrons become less coherent with increasing temperature. To see this, we may write down the dispersions of the hybridization bands as $E_{k,\pm}=1/2[\varepsilon_{k,1}+\varepsilon_{k,2}-i\Gamma\pm((\varepsilon_{k,1}-\varepsilon_{k,2}-i\Gamma)^2+4V^2)^{1/2}]$, where $\varepsilon_{k,1}$ and $\varepsilon_{k,2}$ are the dispersions of the Cu $3d_{xy}$ and Ru $4d$ bands, respectively, $\Gamma$ is the imaginary part of the Cu $3d_{xy}$ self-energy, and $V$ is the effective hybridization. For $\Gamma\ll V$, these give rise to two well-defined hybridization bands. But for $\Gamma\gg V$, we have approximately the expansion, $E_{k,+}=\varepsilon_{k,1}-i(\Gamma-V^2/\Gamma)$ and $E_{k,-}=\varepsilon_{k,2}-iV^2/\Gamma$, such that the two bands are effectively decoupled. Indeed, we see at high temperatures a hole band from the Ru $4d$ orbital and a weak electron band from the Cu $3d$ orbital around the H point. To better compare with experiment, we also plot Fig. 5(d) the density of states near the Fermi energy. We find two peaks below and above the Fermi energy, which shift down with lowering temperature and are in good agreement with the observed two-peak structure in our ARPES data.

Thus we have a good qualitative agreement between the experiment and calculations, which confirms the presence of a flat hybridization band responsible for the heavy electron properties of CCRO. However, unlike the usual Kondo lattice, the Cu $3d$ electrons remain itinerant (albeit very incoherent) at high temperatures. This clearly rules out the Kondo scenario and implies a different mechanism in this compound. Our observation is consistent with the absence of local magnetic moment revealed by NMR[11] and supports the two-band scenario from the moderately correlated electrons (Cu $3d_{xy}$) and itinerant holes (Ru $4d$), leading to an enhanced effective mass in CCRO. The observed peaks in the DOS provide a microscopic support for explaining the maximum feature in the magnetic susceptibility around 200 K.

## CONCLUSIONS

To summarize, we performed ARPES measurements and first-principles calculations on the $d$-electron perovskite heavy-fermion CCRO. The itinerant Cu $3d$ electron band hybridizes with the Ru $4d$ hole band, opens a hybridization gap of $\sim 90$ meV near the Fermi level, and forms flat bands above and below the Fermi level at low temperatures. Our results exclude the Kondo scenario in CCRO, and suggest that some other mechanisms beyond the Kondo scenario such as a two-band model and/or spin-fluctuation may be needed to explain the mechanism of the heavy-fermion CCRO.

## Acknowledgements:

This work was supported by the National Key R&D Program of China (2016YFA0401000，2017YFA0303103), the National Natural Science Foundation of China (11674371, 11974397, 11774401).

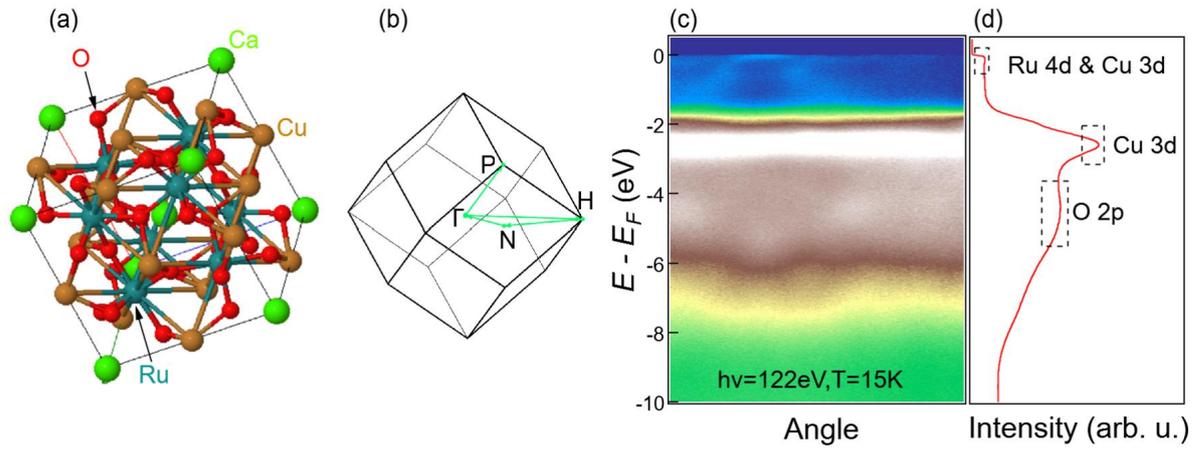

FIG. 1: (Color online) Crystal structure, Brillouin zone (BZ), valence band characterization of $CaCu_3Ru_4O_{12}$. (a) Crystal structure of $CaCu_3Ru_4O_{12}$. (b) First Brillouin zone of body centered cubic (bcc). (c) Valence band spectrum of $CaCu_3Ru_4O_{12}$. (d) Integrated density of states of (c).

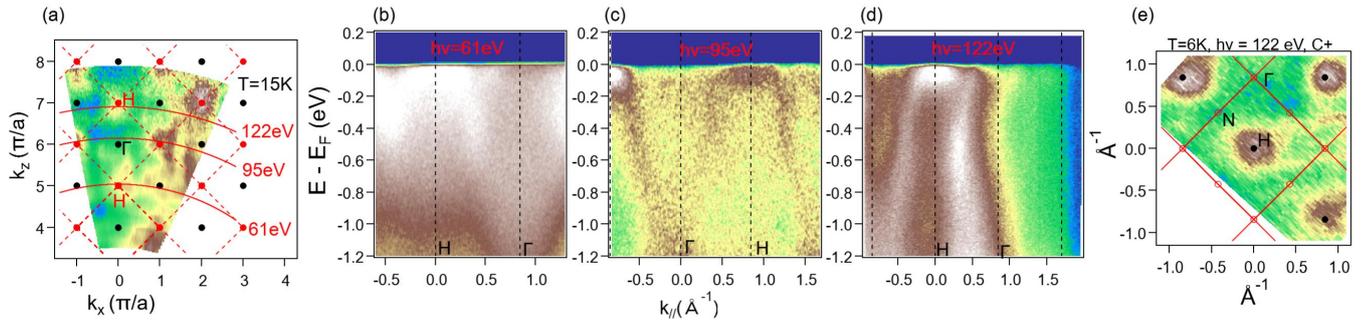

FIG. 2: (Color online) $k_z$ map, several high symmetry cuts and in-plane Fermi surface measured with $hv$ = 122 eV. (a) $k_z$ map obtained from $hv$ = 25 eV to $hv$ = 161 eV in steps of 2 eV at a low temperature of 15 K with circular polarization. Red dashed grids are the brillouin zone boundary of $ac$ plane. $\Gamma$ - H cuts with photon energies at (b) 61 eV, (c) 95 eV, (d) 122 eV. Inner potential $V_0$ = 13 eV was estimated from $k_z$ dispersions. (e) Fermi surface measured with photon energy $hv$ = 122 eV at T = 6 K with circular polarization.

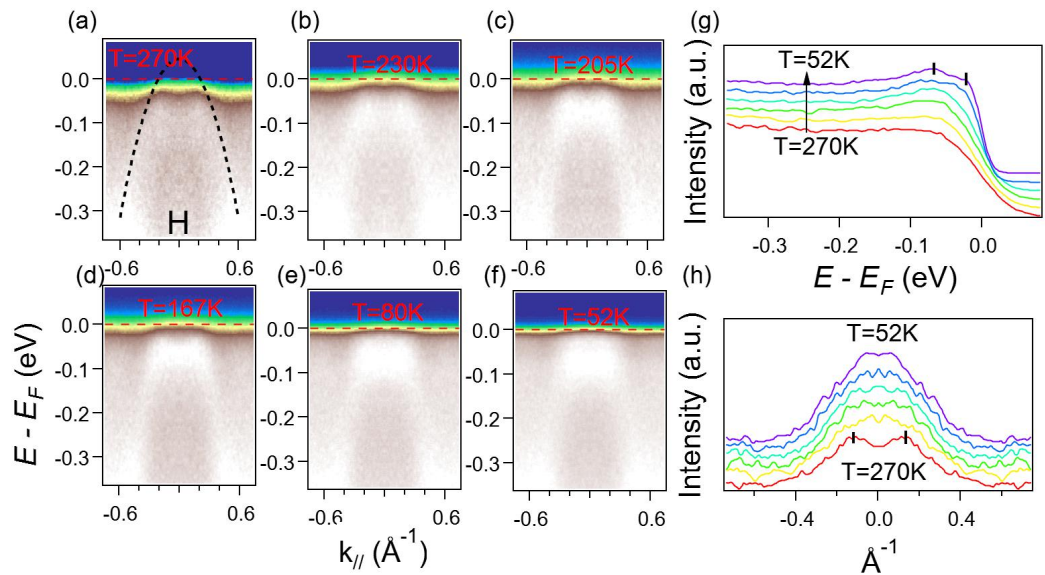

FIG. 3: (Color online) Temperature dependence of band structure around the H point. (a) - (f) Spectra around the H point at corresponding temperatures. (g) EDCs at the H point. (h) MDCs at the Fermi level. The parabolic dashed line in (a) indicates the hole band across the Fermi level at high temperatures. Two short vertical bars in (g) indicate two quasiparticle peaks near the Fermi level at $T = 52$ K. Two short vertical bars in (h) indicate the $k_{FS}$ of hole band at $T = 270$ K.

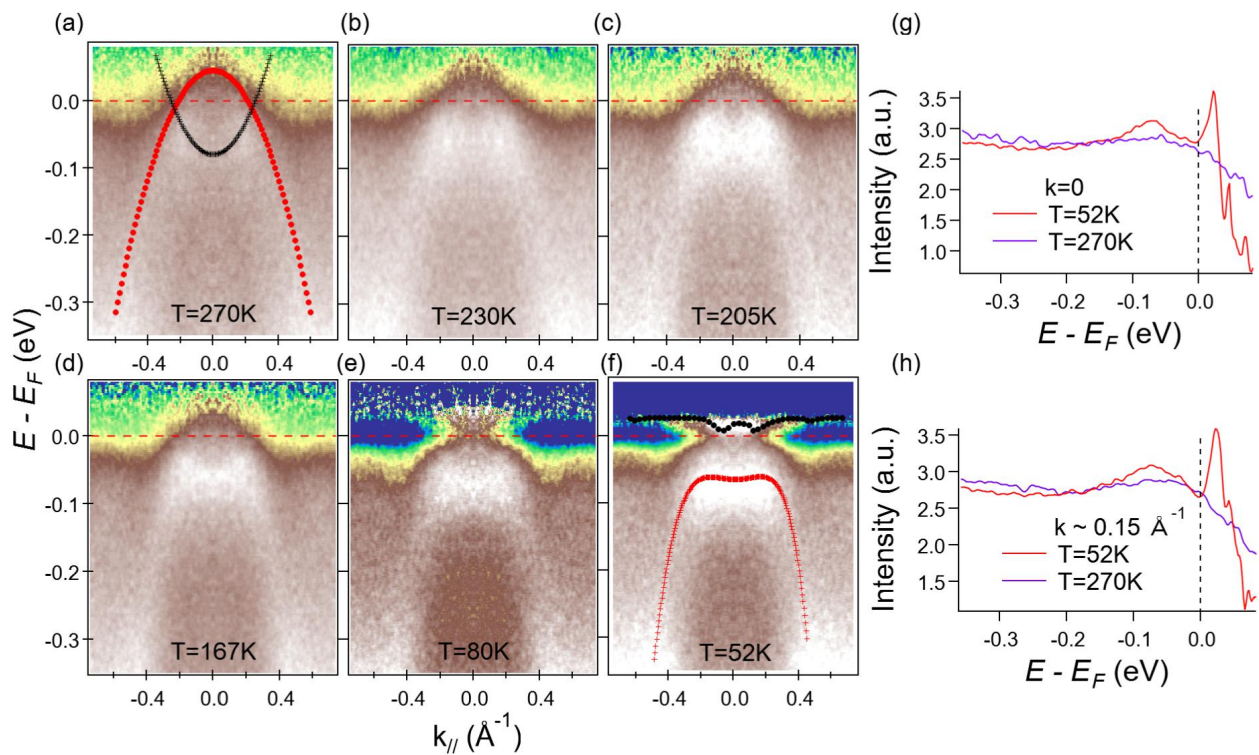

FIG. 4: (Color online) Spectra divided by the resolution-convoluted Fermi-Dirac function to obtain the information above $E_F$ at corresponding temperatures. EDCs at $T$ = 52 K and $T$ = 270 K at (g) $k$ = 0, (h) k $\sim$ 0.15 Å$^{-1}$.

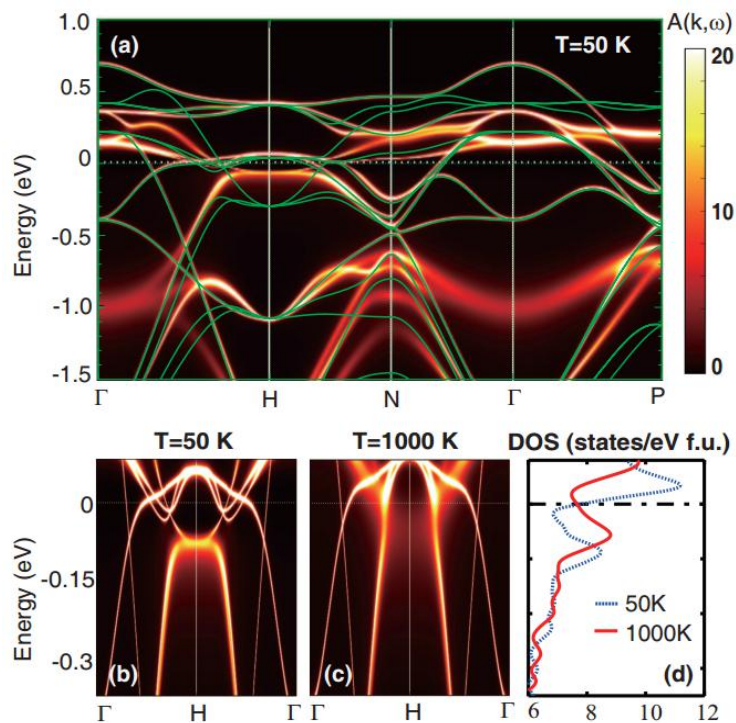

FIG. 5: (Color online) (a) DFT+DMFT spectral functions along Γ- H - N -Γ - P path at 50 K. The DFT bands (green lines) are also plotted for comparison. (b) and (c) Comparison of DFT+DMFT spectral functions at 1000 K and 50 K along Γ - H - Γ path in a smaller energy window. (d) The corresponding density of states at 50 K (blue dashed line) and 1000 K (red solid line), showing a two-peak structure around the Fermi energy and their downshift at lower temperatures.